\documentclass{jkas}

    \def\year{2018} 
    \def\month{July} 
    \def\volume{00} 
    \def\issue{0} 
    \def\beginpage{1} 
    \def\endpage{99} 
    \def\received{} 
    \def\accepted{} 
    \date{Received \received; accepted \accepted}
    
    \usepackage{flushend} 
    \usepackage[colorlinks,
            ps2pdf,
            breaklinks,
            linkcolor=blue,
            urlcolor=blue,
            anchorcolor=blue,
            menucolor=blue,
            citecolor=blue]{hyperref}
    
    \title{
    Source-Frequency Phase-Referencing Observation of AGNs with KaVA Using Simultaneous Dual-Frequency Receiving
    }

    \author[1]{Guang-Yao~Zhao}
    \author[1,2]{Taehyun~Jung}
    \author[1,2,15]{Bong~Won~Sohn}
    \author[3,4]{Motoki~Kino}
    \author[4,5]{Mareki~Honma}
    \author[6]{Richard~Dodson}
    \author[6,7,8]{Mar\'ia~Rioja}
    \author[1]{Seog-Tae~Han}
    \author[4,5]{Katsunori~Shibata}
    \author[1,2,15]{Do-Young~Byun}
    \author[9]{Kazunori~Akiyama}
    \author[10,17]{Juan-Carlos~Algaba}
    \author[11,12]{Tao~An}
    \author[11,13]{Xiaopeng~Cheng}
    \author[1,2]{Ilje~Cho}
    \author[4]{Yuzhu~Cui}
    \author[4,5]{Kazuhiro~Hada}
    \author[1]{Jeffrey~A.~Hodgson}
    \author[11,12]{Wu~Jiang}
    \author[1]{Jee~Won~Lee}
    \author[10]{Jeong~Ae~Lee}
    \author[14]{Kotaro~Niinuma}
    \author[10]{Jong-Ho~Park}
    \author[15]{Hyunwook~Ro}
    \author[16]{Satoko~Sawada-Satoh}
    \author[11,12]{Zhi-Qiang~Shen}
    \author[4]{Fumie~Tazaki}
    \author[10]{Sascha~Trippe}
    \author[1]{Kiyoaki~Wajima}
    \author[11,13]{Yingkang~Zhang}

    \affil[1]{Korea Astronomy and Space Science Institute, Daejeon 34055 Korea; \email{gyzhao@kasi.re.kr}}
    \affil[2]{University of Science and Technology, 217, Gajeong-ro, Yuseong-gu, Daejeon 34113 , Korea;}
    \affil[3]{Kogakuin University, Academic Support Center, 2665-1 Nakano, Hachioji, Tokyo 192-0015, Japan}
    \affil[4]{National Astronomical Observatory of Japan, 2-21-1 Osawa, Mitaka, Tokyo 181-8588, Japan}
    \affil[5]{Department of Astronomical Science, SOKENDAI, 2-21-1, Osawa, Mitaka, Tokyo 181-8588, Japan}
    \affil[6]{ICRAR, M468, University of Western Australia, 35 Stirling  Hwy, Perth 6009, Australia}
    \affil[7]{CSIRO Astronomy and Space Science, 26 Dick Perry Avenue, Kensington WA 6151, Australia}
    \affil[8]{OAN (IGN), Alfonso XII, 3 y 5, 28014 Madrid, Spain}
    \affil[9]{Massachusetts Institute of Technology, Haystack Observatory, 99 Millstone Road, Westford, MA 01886, USA}
    \affil[10]{Department of Physics and Astronomy, Seoul National University, Gwanak-gu, Seoul 08826, Korea}
    \affil[11]{Shanghai Astronomical Observatory, Chinese Academy of Sciences, 80 Nandan Road, Shanghai 200030, China}
    \affil[12]{Key Laboratory of Radio Astronomy, Chinese Academy of Sciences, 210008 Nanjing, People's Republic of China}
    \affil[13]{University of Chinese Academy of Sciences, 19A Yuquanlu, Beijing 100049, People's Republic of China}
    \affil[14]{Graduate School of Sciences and Technology for Innovation, Yamaguchi University, Yoshida 1677-1, Yamaguchi, Yamaguchi 753-8512, Japan}
    \affil[15]{Department of Astronomy, Yonsei University, 134 Shinchon-dong, Seodaemun-gu, Seoul 03722, Korea}
    \affil[16]{Graduate School of Science and Engineering, Kagoshima University, 1-21-35 Korimoto, Kagoshima-shi, Kagoshima 890-0065, Japan}
    \affil[17]{Department of Physics, Faculty of Science, University of Malaya, 50603 Kuala Lumpur, Malaysia}

    \def\corrauthor{
    G.-Y. Zhao
    }

    \def\runningauthor{
    Zhao et al.
    }

    \def\runningtitle{
    Simultaneous Dual-Frequency Observations with KaVA
    }

    \def\keywords{
    galaxies: active --- 
    galaxies: individual: 4C 39.25, 0945+408 --- 
    astrometry ---
    techniques: interferometric
    }

    \def\abstracttext{
    The KVN(Korean VLBI Network)-style simultaneous multi-frequency receiving mode is demonstrated to be promising for mm-VLBI observations. Recently, other Very long baseline interferometry (VLBI) facilities all over the globe start to implement compatible optics systems. Simultaneous dual/multi-frequency VLBI observations at mm wavelengths with international baselines are thus possible. In this paper, we present the results from the first successful simultaneous 22/43 GHz dual-frequency observation with KaVA(KVN and VERA array), including images and astrometric results. Our analysis shows that the newly implemented simultaneous receiving system has brought a significant extension of the coherence time of the 43 GHz visibility phases along the international baselines. The astrometric results obtained with KaVA are consistent with those obtained with the independent analysis of the KVN data. Our results thus confirm the good performance of the simultaneous receiving systems for the non-KVN stations. Future simultaneous observations with more global stations bring even higher sensitivity and micro-arcsecond level astrometric measurements of the targets.
    }

\begin{document}
    \jkashead 


    \section{Introduction}\label{introduction}

Very long baseline interferometry (VLBI) observations at millimeter wavelengths are conventionally difficult due to tight limitations on the integration time imposed by fast phase fluctuations caused by atmospheric propagation effects. The unique quasi-optical (hereafter QO) system implemented on the Korean VLBI Network (KVN) that enables simultaneous multi-frequency receiving overcomes some of these difficulties \citep{han08, han13, jung15}.
 The non-dispersive characteristics of tropospheric propagation effects \citep{rioja11}, which dominate the phase noise at high frequencies, lead to strong correlations between visibility phases at different frequencies \citep[e.g.,][]{jung11}. It is thus possible to calibrate these effects in the high frequency phases with their lower frequency counterparts \citep[e.g.,][]{middelberg05, rioja11, jiang18}. Simultaneous multi-frequency receiving has made this kind of calibration very feasible as no temporal interpolations are necessary \citep[e.g.,][]{rioja11, rioja15, algaba15}. The immediate benefit of applying this method (\emph{so-called frequency phase transfer, FPT}) is the extension of the coherence time at high frequency. \citet{rioja15} found that the coherence time at 130 GHz increased from several seconds to ~20 minutes after FPT. It is possible to further extend it to hours if more than 3 frequencies are simultaneously observed \citep{zhao18}, making the KVN very powerful in detecting weak sources at high frequencies~\citep[e.g.,][]{algaba15}.

Another benefit of the QO system is improved mm-wave astrometry. The long coherence time and weaker dependence on line of sights of the phases after FPT allow switching between sources in order to calibrate residual propagation effects, i.e., source-frequency phase-referencing \citep[SFPR,][]{rioja11}. This will also  provide astrometric information which is a combination of the relative positions between two frequencies in the two sources~\citep[e.g.,][]{rioja11, rioja14, rioja15, jiang18}. For an AGN jet, the frequency-dependent shift in the position of the VLBI core is known as the core-shift \citep{blandford79}. KVN is doing regular astrometric observations with SFPR up to 130~GHz \citep[e.g.,][]{rioja14,rioja15, dodson14, dodson18, yoon18}, and consistent results were found between KVN and the Very Long Baseline Array (VLBA) \citep{rioja14}.

The disadvantages of KVN, however, come mainly from the short baseline lengths and the limited number of baselines. Poor image qualities\citep[e.g.,][]{zhao18} and structural blending effects in astrometric measurements \citep{rioja14, zhao15} are often seen in the results.
Nevertheless, as the simultaneous observing mode has been demonstrated to have great potential for future mm-VLBI observations~\citep[e.g.,][]{dodson17nar}, more stations all over the world begin to implement compatible optical systems.
Non-KVN stations to have been upgraded with QO systems include the 4 VERA (VLBI Exploration of Radio Astrometry) stations in Japan and the Yebes 40m telescope in Spain.  The Australia telescope compact array (ATCA) in Narrabri, Australia can also observe in dual-frequency mode with the KVN by sub-arraying its 6 antennas at different frequencies of 43/86 GHz as was successfully demonstrated in 2014 \citep[e.g.,][]{jung15}. More stations with simultaneous observing capabilities such as the Mopra telescope in Australia will likely become available in the coming few years. The establishment of such a global array will overcome the disadvantages of KVN-only observations.

Test runs of 22/43 GHz simultaneous dual-frequency observations with KaVA (KVN and VERA array) began shortly after the implementation of the QO systems on the VERA stations. A more detailed summary on the commissioning for simultaneous observations with KaVA will be presented in a companion paper \citep{jung18}.

In this paper, we present one of the first simultaneous dual-frequency observations with KaVA. The results indicate good performance of the QO for VERA stations. The paper is structured as follows, details about the observation and data analysis are described in Section~\ref{observation-and-data-analysis}. In Section~\ref{results-and-discussions}, we present results including structural maps, coherence analysis and astrometric measurements, each followed by a brief discussion. We summarize our results in Section~\ref{summary-and-future-aspects}.

    \begin{figure*}[p]
\centering
    \includegraphics[angle=0, width=8.5cm]{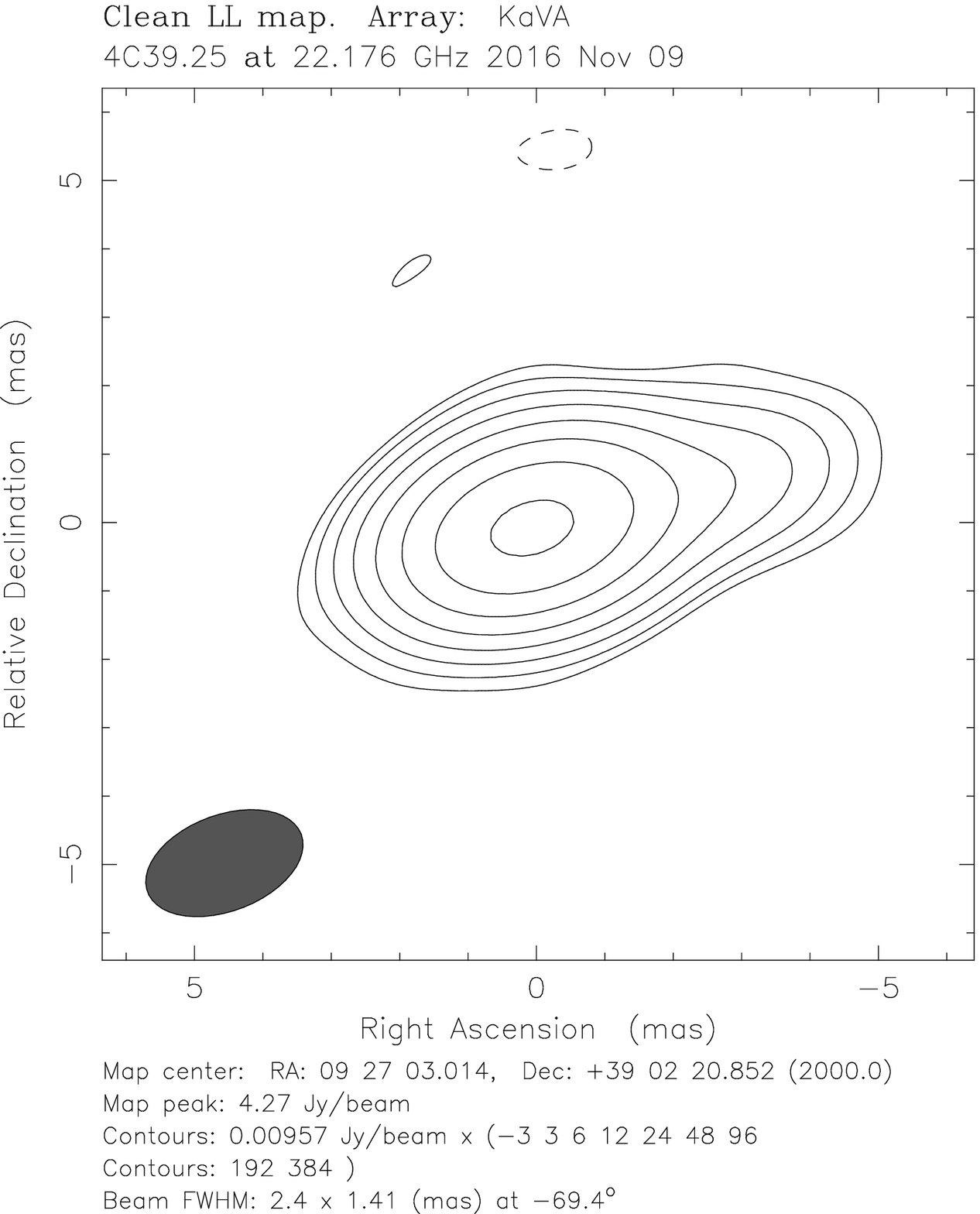}
    \includegraphics[angle=0, width=8.5cm]{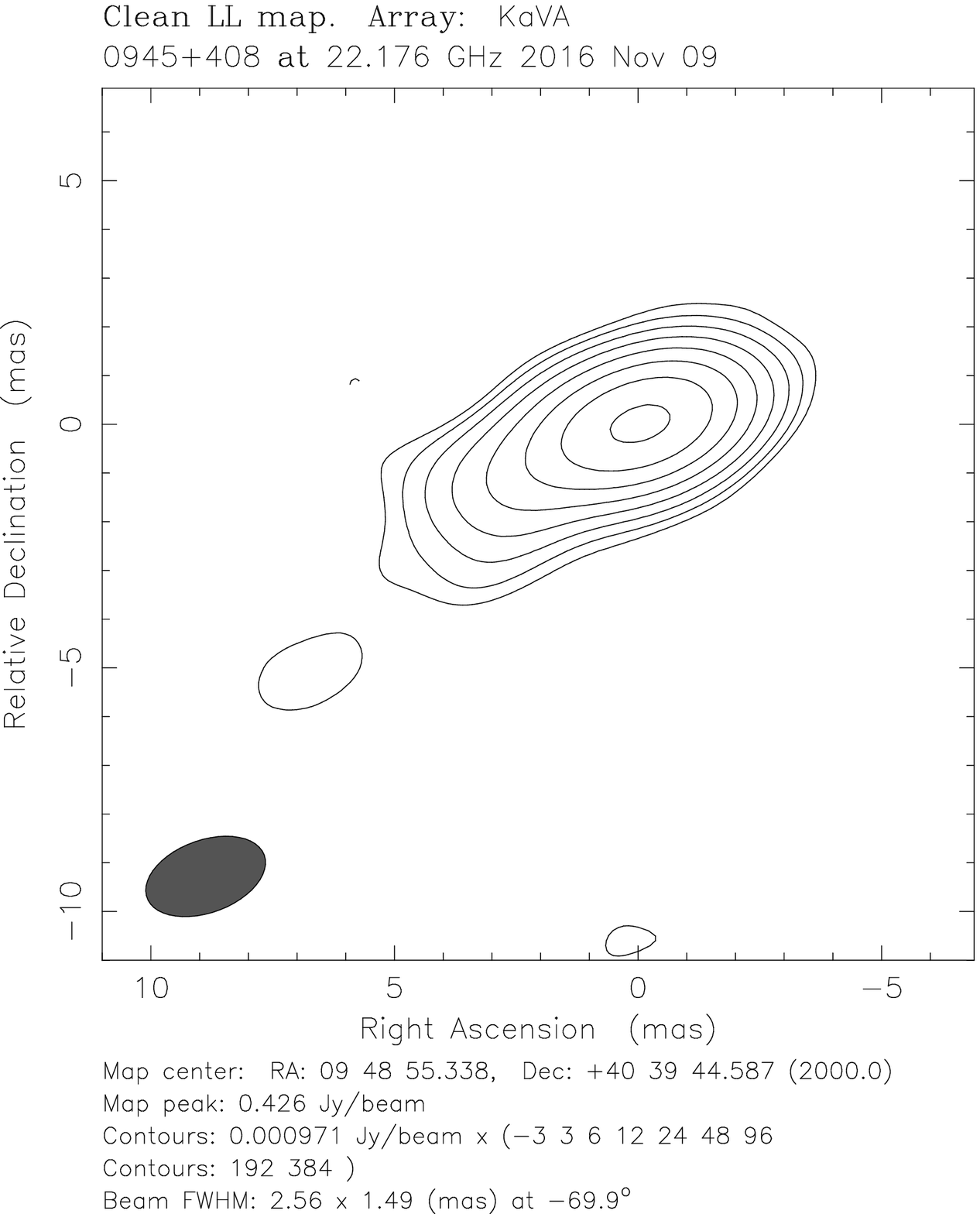} \\ 
    \includegraphics[angle=0, width=8.5cm]{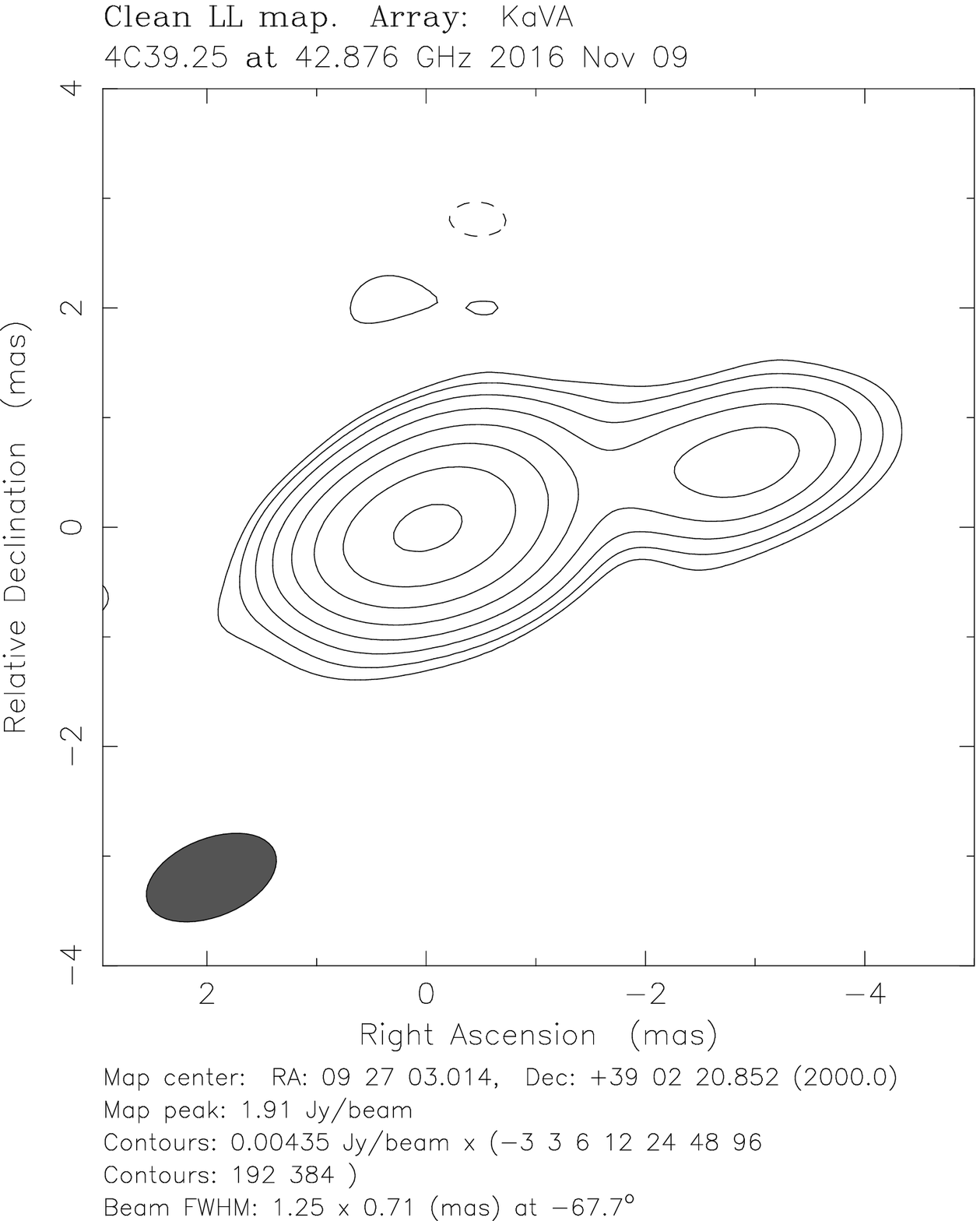}
    \includegraphics[angle=0, width=8.5cm]{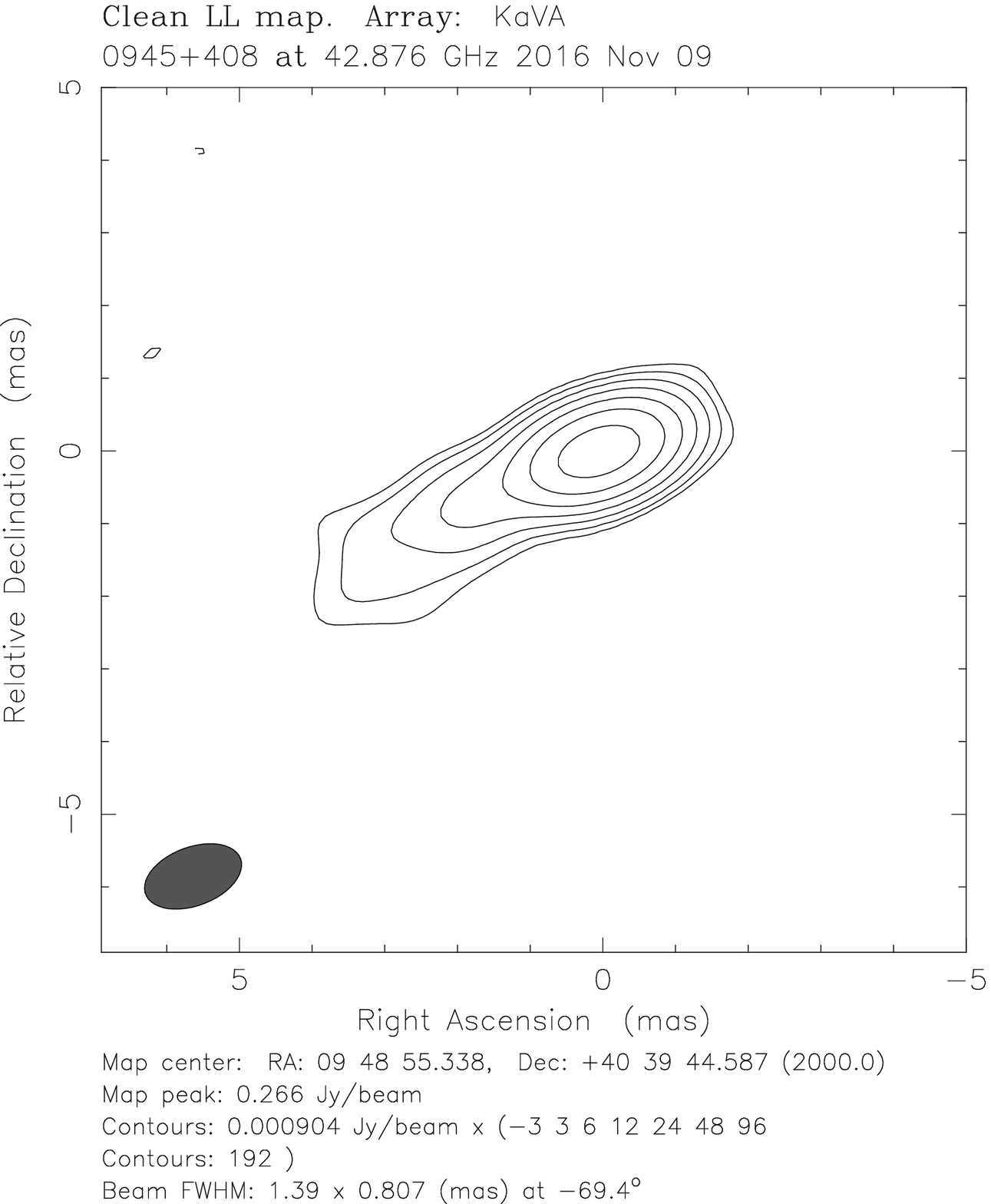}
    \caption{Self-calibrated images of 4C~39.25 (left) and 0945+408 (right) at 22 (upper) and 43 GHz (lower). The beams are shown in the lower left corner. The contour levels start from 3 times the r.m.s level and increase by a factor of 2.}
    \label{fig:maps}
\end{figure*}

    \begin{figure*}[!t]
    \centering
    \includegraphics[trim=3mm 3mm 5mm 10mm, clip, width=15cm]{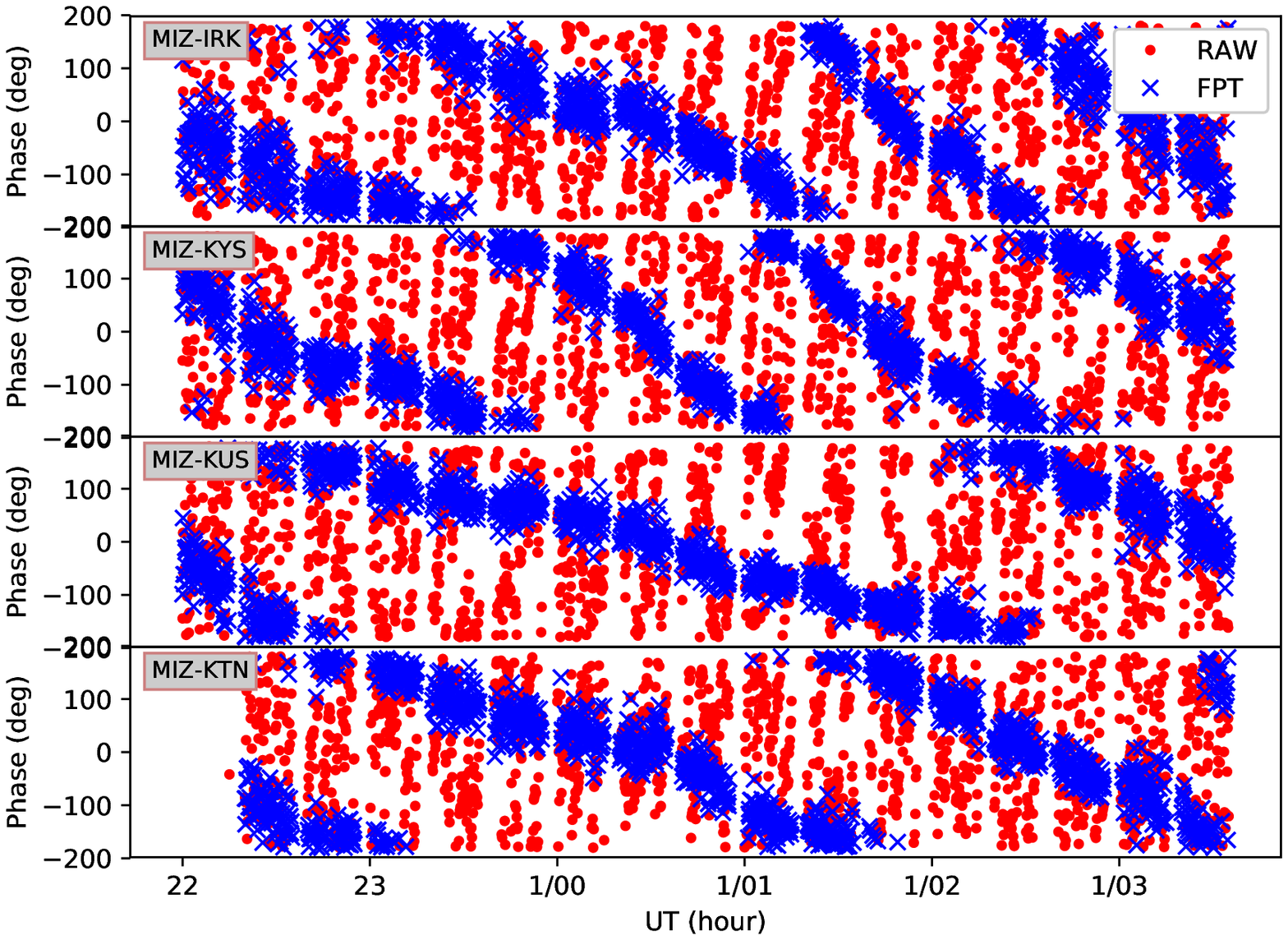}
    \caption{visibility phases of 4C~39.25 along MIZ baselines at 43 GHz before (dots) and after (crosses) applying FPT from 22 GHz; Each point on the plot corresponds to a temporal average of 10 seconds.}
    \label{fig:fpt}
\end{figure*}

\section{Observation and Data Analysis}\label{observation-and-data-analysis}

Simultaneous 22/43 GHz VLBI observations with the joint array of KVN and the international partner stations were successfully carried out in two sessions, on November 7--9, 2016 and March 16--18, 2018, respectively.
The participating antennas included all three KVN telescopes in Korea, the four VERA telescopes in Japan, and the Yebes 40m telescope in Spain. These stations are implemented with KVN compatible dual-frequency QO systems. The data obtained in the second session are yet to be analyzed and the scientific results will be presented in forthcoming papers. Here we summarize the observation and data analysis for 4C~39.25 campaign in the first session.

\subsection{Observation}\label{observation}

The source 4C~39.25 ($RA = 9h27m03.013938s, Dec = +39^\circ 02' 20.851770'', Epoch = J2000$) was observed on November 9, 2016 for 6 hours while the target was setting as seen from East Asian stations. During the first session, fringe-detections with Yebes were not successful and only two VERA stations (Mizusawa and Iriki) were implemented with QO systems.

The observing frequencies were 22.1 and 42.8 GHz. The total recording bandwidth was 256 MHz (128 MHz at each band). The signal was further split into 32 MHz IFs.
The other two VERA stations (Ogasawara and Ishigaki) were observing at 42.8 GHz only and the data were not used in this analysis.
Usually, for dual-frequency calibrations, an integer ratio between frequencies is favored because it is ideal for addressing the \(2n\pi\) phase-wrapping ambiguities \citep[e.g.,][]{middelberg05}.
However, in practice, this requirement cannot always be satisfied, as for example, the rest frequencies of SiO and water maser transitions have a non-integer ratio. In the case of our observation, the frequency windows are limited by the common frequency range among different stations. Nevertheless, FPT with non-integer ratios can be implemented \citep[e.g.,][]{dodson14}.

Most of the observing time was for tracking 4C 39.25 with the aim of checking the system stability over a long time and a wide range of elevations. 4C 39.25 is suitable for this purpose as it passed near the zenith of most of the stations during the observation campaign.
Short (3 minutes) scans on another source, 0945+408 ($RA = 9h48m55.338151s, Dec = +40^\circ 39' 44.586920'', Epoch = J2000$), were inserted every 20 minutes to obtain astrometric measurements using SFPR. The separation between the two sources on the sky plane is \(\sim5\deg\).

Typical weather conditions were found during the observation at most of the sites, with sky opacity less than 0.15 and opacity corrected system temperatures (\(T^{*}_{sys}\)) ranging between 100 and 600 K during most of the observing time, except around UT 0 at Tamna and during the last two hours at Iriki station due to bad weather conditions (i.e., sky opacity reaches nearly 0.3 and $ T^{*}_{sys} >  $1000 K).

After the observations, data recorded at each site were sent to the Korea-Japan joint correlation center (KJCC) and correlated by the hardware correlator \citep{lee15a}. The same correlator is also used for most of the KaVA and EAVN (East Asian VLBI Network) observations.

\subsection{Data Reduction}\label{data-reduction}

The data analysis for the simultaneous 22/43 GHz campaign was carried out with AIPS and Difmap.
First, amplitude calibration was performed in two steps, normalizing the auto-correlation spectra at each correlation interval with AIPS task ACCOR and multiplying the gain curves and the atmospheric opacity corrected system temperatures (\(T^{*}_{sys}\)) with task APCAL.
A correction factor of 1.33 for digital sampling loss and quantization loss was also applied with APCAL \citep{lee15b}.

Phase calibration was conducted in three parallel passes.
First, we performed global fringe-fitting, self-calibration and hybrid mapping for each source at each frequency independently following the standard procedure in regular imaging observations.
The solution intervals for fringe-fitting were set to be the typical coherence times at each frequency, i.e., 60 and 30 seconds for 22 and 43 GHz respectively;
Second, we applied SFPR analysis with the data on all baselines of the five stations.
We performed fringe-fitting at the lower frequency band, 22 GHz, with a solution interval that matches the typical coherence time of the higher frequency (30s) and derived the phase solutions for the higher frequency band data that are obtained at the same time.
We first scaled up the phase solution for the first IF (reference frequency, 22.1 GHz) with a factor of 2 and then calculated the phase solutions for each IF (\(\phi_{IF}\)) with the multi-band delay (\(\tau_{MB}\)) and the frequency difference (i.e., \(\phi_{IF} = 2~\times~\phi_{22.1} + \tau_{MB}~\times~(\nu_{IF}-2~\times~22.1)\), where \(\phi_{22.1}\) and \(\nu_{IF}\) stands for the phase solutions at 22.1 GHz and the frequency of each IF, respectively).
After applying the phase solutions derived from the lower frequency (FPT), we estimated the coherence time at the higher frequency band using the simple approach demonstrated in \citet{rioja12}. 
We then ran a second fringe-fitting on the 4C 39.25 data at the higher frequency and applied the solutions to both sources by cubic temporal interpolation.
Finally, we imaged the 0945+408 data and measured the position offset from the phase center which corresponds to the combination of the relative position between two frequencies of the two sources, i.e., the core-shift;
In the third pass, we followed the same procedure as in the second one but with KVN data alone to provide a consistency-check on the reliability of the results.

\begin{figure}[!t]
    \centering
    \includegraphics[trim=4mm 1mm 18mm 15mm, clip, width=84mm]{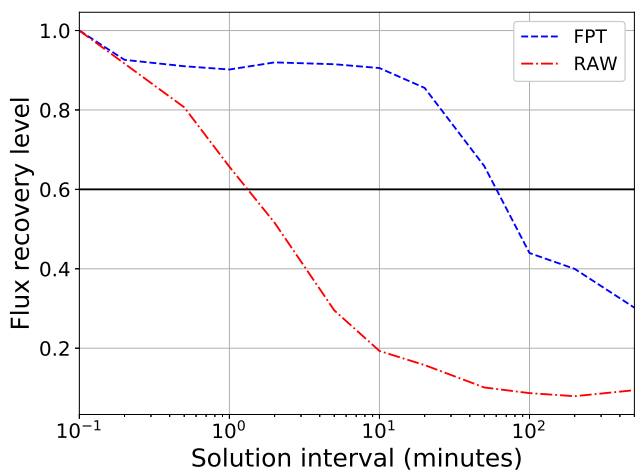}
    \caption{Fractional flux recovery as functions of solution intervals for phase self-calibration, for the visibility phases before (red) and after (blue) applying FPT. The coherence time corresponds to a fractional flux recovery of ~0.6.}
    \label{fig:coher}
\end{figure}

\section{Results and Discussions}\label{results-and-discussions}

In this section, we present the results of 4C 39.25 \(\&\) 0945+408 observations during the first international simultaneous 22/43 GHz dual-frequency campaign.
The main results include the amplitude self-calibrated dual-frequency maps of both sources, the extension of the coherence time at the higher frequency after applying FPT, and the astrometric measurements with SFPR which is the combined core-shift (or \emph{centroid-shift}) of the two sources.
Each of these results are followed by a brief comparison with either previous KaVA imaging results or the KVN FPT and astrometric results.


\subsection{Hybrid Maps}\label{hybrid-maps}

KaVA is now conducting imaging observations at 22 and 43 GHz regularly (not simultaneously) for over 1000 hours per year with several large programs running~\cite[e.g.,][]{kino15}.
For projects that require dual frequency images, observations at different frequencies are usually done on two successive days.
The imaging capability of KaVA was evaluated in details by \citet{niinuma14}.
In this section, we compare the morphology with previous observations by KaVA to check the reliability of the system.
Detailed studies of source structure and spectra will be presented together with the result from the second session and follow-up observations in forthcoming papers.

The self-calibrated hybrid maps of the two sources at two frequencies are shown in Figure~\ref{fig:maps}.
Our observation reveals similar structures of the targets as compared with previous works.
For 4C 39.25, our maps show the peculiar core-jet structure of this source, with a bright, complex, optically-thin feature and a relatively dimmer, compact core component at a distance of $\sim3~~\mbox{mas}$ in the west from the brightest feature.
This morphology is consistent with previous studies~\citep[e.g.,][]{niinuma14}.
However, the resolution in our 22~GHz map is not sufficient to fully resolve the structure of this source as can be seen from the map.
For 0945+408, we find a one-sided core-jet structure with the jet extending in the southeast direction. A slight change of the jet axis direction is shown at $\sim2~~\mbox{mas}$ from the core. This morphology is similar as presented elsewhere, for instance in the 15~GHz results by MOJAVE project~\citep{lister13}. The weak, extended emissions at $\sim7~~\mbox{mas}$ and beyond from the core revealed by MOJAVE are not obvious on our maps, suggesting that these diffuse emissions have steep spectra.

\begin{figure*}[t!]
  \centering
    \includegraphics[width=8.5cm]{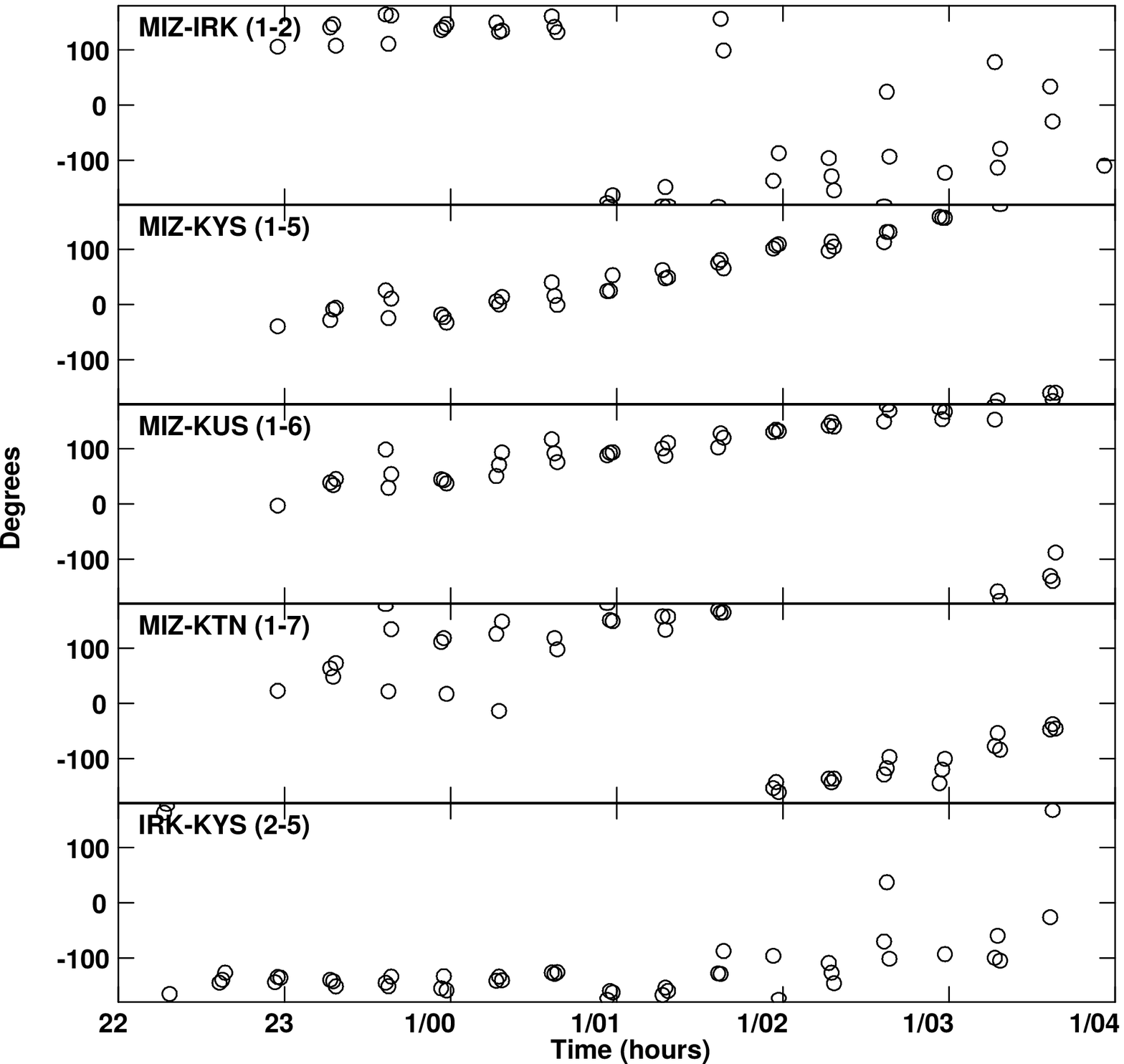}
    \includegraphics[width=8.5cm]{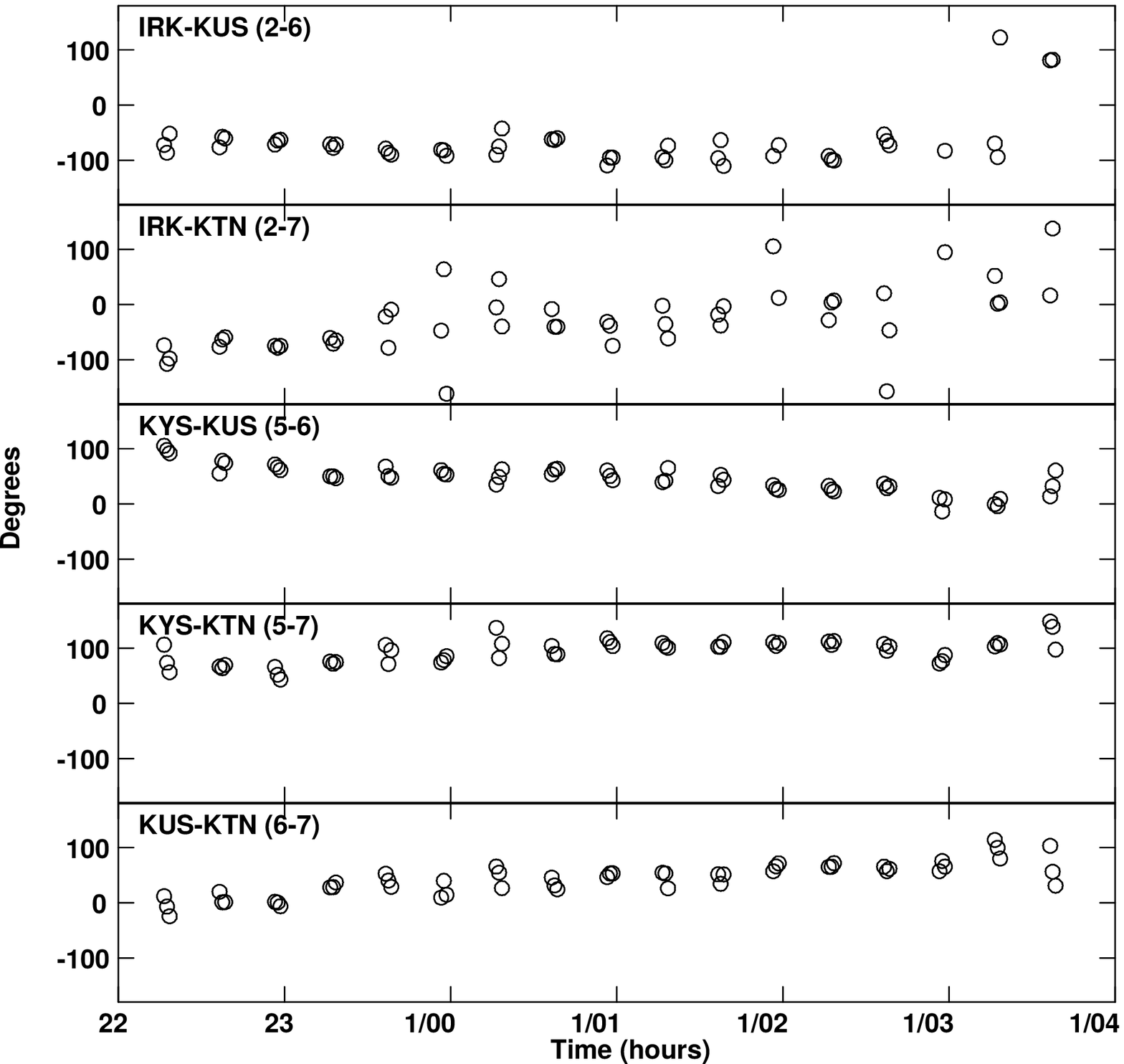}
    \caption{SFPRed visibility phases of 0945+408 on all baselines at 43 GHz with a temporal average of 60s. The numbers in the parentheses in each panel correspond to the station codes of each antenna used by the correlator.}
    \label{fig:vplot_sfpr}
\end{figure*}

\subsection{Extension of Coherence Times}\label{extension-of-coherence-times}

One of the most important improvements to come from implementing the QO system is the extension of the coherence time at higher frequencies by applying FPT \citep[e.g.,][]{rioja15, algaba15, dodson17nar}.
In the second pass of our data analysis, we calibrate the higher frequency data with the fringe solutions obtained at the lower frequency.
The 43 GHz visibility phases of 4C 39.25 at Mizusawa baselines before and after this calibration are
shown in Figure \ref{fig:fpt}.
This figure clearly demonstrates the power of the simultaneous dual-/multi-frequency receiving.
As we can see from the plots, before applying FPT, which is equivalent to a single frequency observation, the phases are dominated by propagation errors. The phase rotation can be much larger than 1 radian within several minutes.
After applying FPT, the phases are significantly better aligned. The time taken for the phase to rotate around 1 radian approaches several tens of minutes. Such an improvement in the coherence is comparable with those seen in the KVN-only experiments~\citep[e.g.,][]{rioja14, jung15}

In order to quantify the extension of coherence time, we followed a simple procedure presented in \citet{rioja12}.
We loaded the raw and FPTed data into Difmap and performed self-calibration over a series of integration times.
We then imaged the calibrated data and compared the peak flux densities on the map with the one produced with the shortest solution interval, which is 0.1 minutes, much shorter than the typical coherence time at this frequency.
In figure \ref{fig:coher} we show the percentage of the recovered peak flux density compared with the latter as a function of the integration time for each dataset.
The coherence time is defined as the integration time with which the flux recovery level is 0.6~\citep{rioja12,rioja15}.
We found the coherence time at 43 GHz is 1.5 and 60 minutes before and after applying FPT, respectively.

The limitation of coherence time after FPT comes mainly from ionospheric and instrumental propagation effects.
These effects are usually slow varying compared with their tropospheric counterparts as can be seen from the figures.
With KVN 4-frequency observations, \citet{rioja15} reported the coherence time was extended to 20 minutes up to 130 GHz after FPT.
At 43 GHz, this value is expected to be even longer because of the smaller frequency ratio and more stable instrumental performance.
Our KaVA results are quite consistent with such an expectation for KVN which confirms the performance of the QO system for the VERA stations.

We note this is only a simple estimation of the overall coherence time of the observations.
A more detailed analysis on the phase coherence at each baseline using Alan standard deviation and coherence function will be presented in a companion paper \citep[][]{jung18}.

    \subsection{Astrometric Measurements}\label{astrometric-measurements}

\begin{figure}[!t]
    \centering
    \includegraphics[trim=0mm 0mm 0mm 5mm, clip, width=8.4cm]{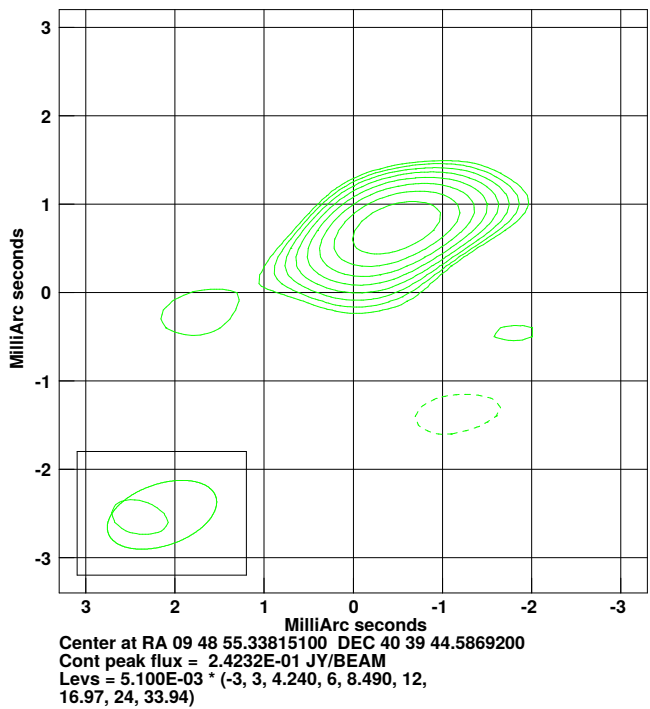}
    \caption{KaVA SFPRed map of 0945+408 at 43 GHz. The beam is $1.28~\times~0.70~\mbox{mas}$ with a position angle of $-71^{\circ}$. The grid serves as a visual guide for the offset of the peak of brightness from the center of the map.}
    \label{fig:map_sfpr}
\end{figure}

In this section, we present the astrometric results by applying the calibrations of 4C 39.25 FPTed data to those of 0945+408, which correspond to the second pass of our data analysis.
An example of the SFPRed phases of 0945+408 along all baselines are shown in Figure \ref{fig:vplot_sfpr} and a corresponding map of the source is shown in Figure \ref{fig:map_sfpr}.
Point source models for fringe-fitting were used in this case.
Compared with the self-calibrated map in Figure~\ref{fig:maps}, although the noise level in Figure~\ref{fig:map_sfpr} is higher, astrometric information of the target is preserved in this map as an offset between the peak position and the map center can be seen in the figure.
This offset corresponds to the combined core-shift or centroid-shift of the two sources.

\begin{figure}[!t]
    \centering
    \includegraphics[trim=10mm 10mm 19mm 24mm, clip, width=8.4cm]{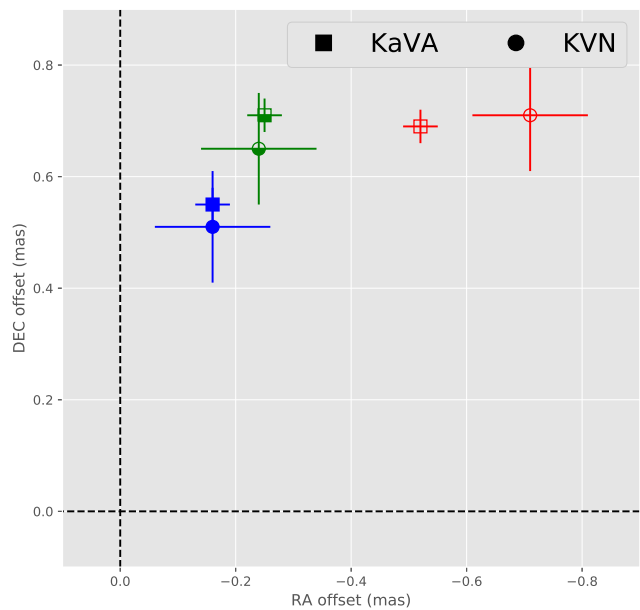}
    \caption{Comparison of astrometric results from the SFPR analysis on the KaVA (squares) and KVN (circles) data with point source models (open symbols), structural models in Section \ref{hybrid-maps} (half-filled symbols), and structual models from Niinuma et al. 2014 for 4C~39.25 (filled symbols).}
    \label{fig:astrometry}
\end{figure}

The reliability of KVN astrometric performance was confirmed in \citet{rioja14} by comparing with the VLBA results.
We compare our results with those obtained from independent parallel analysis on the KVN data only which corresponds to the third pass of our data analysis.
We perform our comparison in three different cases with different structure models used for fringe-fitting, 1) with point source models, 2) with the structural models from the imaging analysis on the data, i.e., those in Section \ref{hybrid-maps}, and 3) with the models from \citet{niinuma14} for 4C 39.25 which were obtained with the full KaVA array but the same as case 2) for 0945+408.
The three pairs of results are plotted in Figure~\ref{fig:astrometry}.
We note that although the shift is relatively large, which indicates the presence of structural blending effects, the measurements of KaVA and KVN agree well with each other, \(2~\sigma\) for point source model and less than \(1~\sigma\) when structure models are used for fringe-fitting.
The results thus confirm the reliability of the performance of QO systems for VERA stations.
Furthermore, because KaVA has higher angular resolutions and image sensitivities, the accuracy of the measurements (\emph{beamsize divided by signal-to-noise ratio}) is much higher than those of KVN. For instance, in Figure~\ref{fig:map_sfpr}, the major axis size of the synthesized beam is $1.3~~\mbox{mas}$ and the dynamic range is about $47$, which result in an astrometric accuracy of $28~~\mbox{\(\mu\)as}$, while for the case with KVN alone, the corresponding values are $3.1~~\mbox{mas}$, $32$, and $96~~\mbox{\(\mu\)as}$, respectively.

Structural blending effects are often seen in KVN astrometric results~\citep[e.g.,][]{rioja14}.
Although our KaVA observation has higher angular resolution, we still find a large position offset on the SFPRed map.
This offset is $\sim1/4$ of the 22~GHz beamsize and along the same direction of the major axis of the beam when point source models are used. When the structural models are used for the SFPR analysis, the amount of offset gets smaller (Figure~\ref{fig:astrometry}).
The most likely reason for this large structure blending effect is the low resolution at 22 GHz as described in Section~\ref{hybrid-maps}. Furthermore, as Iriki station is located close to the Korean peninsula, in the current array configuration, Mizusawa is the only station that has long baselines ($>$ 1000 km).
This means in each triangle of the closure phases, there is at least one short baseline, which is comparable in length to a KVN baseline.
This explains the comparable level of structural blending effect because the standard VLBI data analysis utilize the closure phase during global fringe-fitting and self-calibration.

The limitations of the array with the fact that all the long baselines are associated with one single station will be overcome when simultaneous dual-frequency observations are available with the full KaVA array. The participation of Ogasawara and Ishigaki stations will not only double the longest baseline length, but also will provide closure triangles with all baselines longer than 1200 km (e.g., Mizusawa-Ishigaki-Yonsei triangle). From Figure~\ref{fig:astrometry}, we can notice that when models obtained with the full KaVA array for 4C 39.25 \citep{niinuma14} are used, the offset becomes much smaller than without. If we were to use full-array models for both sources, the offset should become more reasonable. The second observing session performed in March 2018 and future upcoming observations will be able to provide valuable scientific results.

    \section{Summary and Future Aspects}\label{summary-and-future-aspects}

We have carried out simultaneous 22/43 GHz dual frequency observations with the KaVA array.
Our study has confirmed the performance of the quasi-optics systems implemented on VERA stations, as the results:
\begin{itemize}
    \item revealed similar structures of the target sources compared with previous single frequency imaging observations.
    \item confirmed the significant extension of the coherence time at the higher frequency from $\sim$1 minute to several tens of minutes by applying FPT.
    \item found consistent results between the astrometric measurements obtained with KaVA and KVN.
\end{itemize}
Future observations with the full KaVA array and more stations (e.g., in Spain and Australia) will further improve the resolution and sensitivity of the images as well as more accurate astrometric measurements.
The dual-frequency simultaneous observing mode will save approximately half of the time for the dual frequency observations.
Running large programs with simultaneous observations will bring new scientific outcomes (e.g., astrometry) without losing current objectives with more machine time becoming available for new scientific projects, especially as more stations join the array.

\acknowledgments

We are grateful to all staff members at the related stations and KJCC who helped to install the QO systems, to operate the array and to correlate the data.
The KVN is a facility operated by the Korea Astronomy and Space Science Institute.
VERA is a facility operated by the National Astronomical Observatory of Japan in collaboration with associated universities in Japan.
G.-Y. Zhao and T. Jung are supported by Korea Research Fellowship Program through the National Research Foundation of Korea (NRF) funded by the Ministry of Science and ICT (NRF-2015H1D3A1066561). J. C. Algaba acknowledges support from the National Research Foundation of Korea (NRF) via grant NRF-2015R1D1A1A01056807. J. Hodgson is supported by Korea Research Fellowship Program through the National Research Foundation of Korea(NRF) funded by the Ministry of Science and ICT(2018H1D3A1A02032824).


\begin{thebibliography}{}

\bibitem[Algaba et al.(2015)]{algaba15} Algaba, J.-C., Zhao, G.-Y., Lee, S.-S., et al.\ 2015, Interferometric Monitoring of Gamma-Ray Bright Active Galactic Nuclei II: Frequency Phase Transfer, JKAS, 48, 237
\bibitem[Beasley \& Conway(1995)]{beasley95} Beasley, A.~J., \& Conway, J.~E.\ 1995, VLBI Phase-Referencing, Very Long Baseline Interferometry and the VLBA, 82, 327
\bibitem[Blandford \& K{\"o}nigl(1979)]{blandford79} Blandford, R.~D., \& K{\"o}nigl, A.\ 1979, Relativistic Jets as Compact Radio Sources, ApJ, 232, 34
\bibitem[Dodson et al.(2014)]{dodson14} Dodson, R., Rioja, M.~J., Jung, T., et al.\ 2014, Astrometrically Registered Simultaneous Observations of the 22 GHz H$_{2}$O and 43 GHz SiO Masers toward R Leonis Minoris Using KVN and Source/Frequency Phase Referencing, AJ, 148, 97
\bibitem[Dodson et al.(2017)]{dodson17nar} Dodson, R., Rioja, M.~J., Jung, T., et al.\ 2017, The Science Case for Simultaneous mm-Wavelength Receivers in Radio Astronomy, NewAR, 79, 85
\bibitem[Dodson et al.(2018)]{dodson18} Dodson, R., Rioja, M., Bujarrabal, V., et al.\ 2018, Registration of H$_{2}$O and SiO Masers in the Calabash Nebula to Confirm the Planetary Nebula Paradigm, MNRAS, 476, 520
\bibitem[Han et al.(2008)]{han08}Han, S.-T., Lee, J.-W., Kang, J., et al. 2008, Millimeter-Wave Receiver Optics for Korean VLBI Network, IJIMW, 29, 69
\bibitem[Han et al.(2013)]{han13}Han, S.-T., Lee, J.-W., Kang, J., et al. 2013, Korean VLBI Network Receiver Optics for Simultaneous Multifrequency Observation: Evaluation, PASP, 125, 539
\bibitem[Jiang et al.(2018)]{jiang18} Jiang, W., Shen, Z., Jiang, D., et al.\ 2018, VLBI Imaging of M81* at $\lambda = 3.4$ mm with Source-Frequency Phase-Referencing, ApJL, 853, 14
\bibitem[Jung et al.(2011)]{jung11} Jung, T., Sohn, B.~W., Kobayashi, H., et al.\ 2011, First Simultaneous Dual-Frequency Phase Referencing VLBI Observation with VERA, PASJ, 63, 375
\bibitem[Jung et al.(2015)]{jung15} Jung, T., Dodson, R., Han, S.-T., et al.\ 2015, Measuring the Core Shift Effect in AGN Jets with the Extended Korean VLBI Network, JKAS, 48, 277
\bibitem[Jung et al.(2019)]{jung18} Jung, T., et al.\ 2019, JKAS, in preperation.
\bibitem[Kino et al.(2015)]{kino15} Kino, M., Niinuma, K., Zhao, G.-Y., et al.\ 2015,
Key Science Observations of AGNs with KaVA Array, PKAS, 30, 633
\bibitem[Lee et al.(2015a)]{lee15a}Lee, S.-S., Oh, C. S., Roh, D. G., et al. 2015a, A New Hardware Correlator in Korea: Performance Evaluation Using KVN Observations, JKAS, 48, 125
\bibitem[Lee et al.(2015b)]{lee15b} Lee, S.-S., Byun, D.-Y., Oh, C.~S., et al.\ 2015b, Amplitude Correction Factors of Korean VLBI Network Observations, JKAS, 48, 229
\bibitem[Lister et al.(2013)]{lister13} Lister, M.~L., Aller, M.~F., Aller, H.~D., et al.\ 2013, MOJAVE. X. Parsec-Scale Jet Orientation Variations and Superluminal Motion in Active Galactic Nuclei, AJ, 146, 120
\bibitem[Middelberg et al.(2005)]{middelberg05}  Middelberg, E., Roy, A.~L., Walker, R.~C., \& Falcke, H.\ 2005, VLBI Observations of Weak Sources Using Fast Frequency Switching, A\&A, 433, 897
\bibitem[Niinuma et al.(2014)]{niinuma14}Niinuma, K., Lee, S.-S., Kino, M., et al.\ 2014, VLBI Observations of Bright AGN Jets with the KVN and VERA Array (KaVA): Evaluation of Imaging Capability, PASJ 66, 103
\bibitem[Rioja \& Dodson(2011)]{rioja11}Rioja, M., \& Dodson, R. 2011, High-Precision Astrometric Millimeter Very Long Baseline Interferometry Using a New Method for Atmospheric Calibration, AJ, 141, 114
\bibitem[Rioja et al.(2012)]{rioja12} Rioja, M., Dodson, R., Asaki, Y., et al.\ 2012, The Impact of Frequency Standards on Coherence in VLBI at the Highest Frequencies, AJ, 144, 121
\bibitem[Rioja et al.(2014)]{rioja14}Rioja, M., Dodson, R., Jung, T. H., et al., 2014, Verification of the Astrometric Performance of the Korean VLBI Network, Using Comparative SFPR Studies with the VLBA at 14/7 mm, AJ, 148, 84
\bibitem[Rioja et al.(2015)]{rioja15} Rioja, M.~J., Dodson, R., Jung, T., \& Sohn, B.~W.\ 2015, The Power of Simultaneous Multifrequency Observations for mm-VLBI: Astrometry up to 130 GHz with the KVN, AJ, 150, 202
\bibitem[Yoon et al.(2018)]{yoon18} Yoon, D.-H., Cho, S.-H., Yun, Y., et al.\ 2018, Astrometrically Registered Maps of H$_{2}$O and SiO Masers toward VX Sagittarii, Nature Communications, 9, 2534
\bibitem[Zhao et al.(2018)]{zhao18} Zhao, G.-Y., Algaba, J.~C., Lee, S.~S., et al.\ 2018, The Power of Simultaneous Multi-Frequency Observations for mm-VLBI: Beyond Frequency Phase Transfer, AJ, 155, 26
\bibitem[Zhao et al.(2015)]{zhao15} Zhao, G.-Y., Jung, T., Dodson, R., Rioja, M., \& Sohn, B.~W.\ 2015, KVN Source-Frequency Phase-Referencing Observation of 3c 66A and 3c 66B, PKAS, 30, 629
\end{thebibliography}
\end{document}